\documentclass[twocolumn,prl,showpacs,amsmath,amssymb]{revtex4}
\usepackage{graphicx}
\usepackage{dcolumn}
\usepackage{bm}

\begin{document}

\title{Explosive fragmentation of thin ceramic tube using pulsed power}

\author{Hiroaki Katsuragi}
\email{katsurag@asem.kyushu-u.ac.jp}
\affiliation{Department of Applied Science for Electronics and Materials, Interdisciplinary Graduate School of Engineering Sciences, 
Kyushu University, 6-1 Kasugakoen, Kasuga, 
Fukuoka 816-8580, Japan}

\author{Satoshi Ihara}
\affiliation{Department of Electrical and Electronic Engineering, Faculty of Science and Engineering, Saga University, 1 Honjo-machi, Saga 840-8502, Japan}

\author{Haruo Honjo}
\affiliation{Department of Applied Science for Electronics and Materials, Interdisciplinary Graduate School of Engineering Sciences, 
Kyushu University, 6-1 Kasugakoen, Kasuga, 
Fukuoka 816-8580, Japan}

\date{\today}

\begin{abstract}
This study experimentally examined the explosive fragmentation of thin ceramic tubes using pulsed power. A thin ceramic tube was threaded on a thin copper wire, and high voltage was applied to the wire using a pulsed power generator. This melted the wire and the resulting vapor put pressure on the ceramic tube, causing it to fragment. We examined the statistical properties of the fragment mass distribution. The cumulative fragment mass distribution obeyed the double exponential or power-law with exponential decay. Both distributions agreed well with the experimental data. We also found that the weighted mean fragment mass was scaled by the multiplicity. This result was similar to impact fragmentation, except for the crossover point. Finally, we obtained universal scaling for fragmentation, which is applicable to both impact and explosive fragmentation.

\end{abstract}


\maketitle

Fragmentation is one of the most ubiquitous phenomena in nature \cite{Beysens1}. Well known examples include nuclear collision, the impact of brittle materials, and meteorites formed from asteroids. Fragmentation phenomena occur at all scales, from micro to macro, and most occur very rapidly. Consequently, it is difficult to observe fragmentation phenomena directly. The statistical properties of the fragment mass (or size) distribution have been investigated frequently, and a scaling method is very useful for the analysis. This method is also used to study earthquakes, turbulence, and many other non-equilibrium phenomena. Since the scaling laws are universal in many cases, we expect analogies across such phenomena. In particular, brittle fragmentation is a very simple example. Therefore, we believe that investigation of brittle fragmentation is indispensable for understanding non-equilibrium phenomena.

Fragment mass distribution obeys the power-law form $N(m) =\int^{\infty}_{m}n(m')dm' \sim m^{1-\tau}$, that is, $n(m) \sim m^{-\tau}$, where $m$, $n(m)$, and $\tau$ are the fragment mass, fragment number of mass $m$, and characteristic exponent, respectively. Simple experiments examining brittle fragmentation have indicated the universality of fragmentation phenomena with this type of distribution \cite{Ishii1,Oddershede1,Kadono1}. Subsequently, numerical simulations have been performed \cite{Hayakawa1,Kun1,Astrom2,Astrom3,Astrom4,Diehl1,Behera1}. Recently, {\AA}str\"om et al. proposed a generic scaling form, which consists of a combination of the power-law together with exponential decay and exponential functions \cite{Astrom4}. This fits many classes of experiments and simulations, because it contains several typical function forms as limiting cases.

Kun and Herrmann studied the transition between the damaged and fragmented states \cite{Kun1}. This problem is the key to understanding the universality of fragmentation. They proposed that the transition involves the universality of percolation. By contrast, {\AA}str\"om et al. considered another critical behavior in the fragmentation transition \cite{Astrom2}. Katsuragi et al. carried out experiments to check these predictions \cite{Katsuragi1,Katsuragi2}. They also revealed the role of the log-normal distribution. A binomial multiplicative model was able to explain the log-normal distribution and multi-scaling nature of impact fragmentation.

Through these simulations and experiments, the outline of brittle fragmentation has been clarified systematically; however, many unsolved problems remain. One that has received recent attention is the fragmentation of shells \cite{Wittel1}. Wittel et al. examined impact and explosive fragmentation of eggshells in both simulations and experiments. They reported that explosive fragmentation exhibits an abrupt transition, whereas impact fragmentation shows a continuous transition. These results suggest that there is a new kind of universality that reveals the explosive fragmentation of shells. In addition, a double exponential (not the power-law) fragment mass distribution has been observed in explosive fragmentation experiments \cite{Mock1} and simulations \cite{Holian1}. These results indicate the unresolved features of explosive fragmentation. Therefore, more investigations of explosive fragmentation are needed to consider fragmentation universality.

In general, explosive fragmentation experiments are difficult to conduct because of the danger of the explosion. We performed relatively safe explosive fragmentation using a pulsed power generator. In this Letter, we report the results of our experiments. Moreover, we compare the results with previous impact fragmentation experiments. Finally, we discuss the universal scaling of impact and explosive fragmentation.

\begin{figure}
\scalebox{0.5}[0.5]{\includegraphics{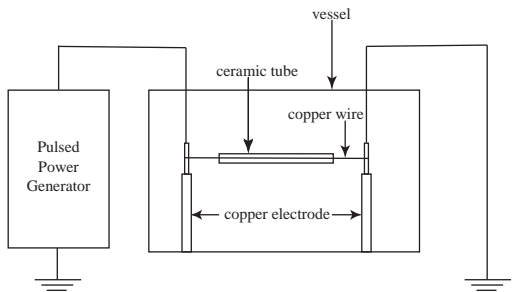}}
\caption{Schematic illustration of the experimental apparatus. A ceramic tube was threaded on a copper wire, and a pulsed high voltage was applied by the pulsed power generator.}
\label{fig:apparatus}
\end{figure}

We used a pulsed power generator consisting of four capacitors with a capacitance of $0.1$ ${\mu}$F each, i.e., $0.4$ $\mu$F in total \cite{Ihara1}. Figure \ref{fig:apparatus} shows a schematic image of the experimental apparatus. The capacitors were charged to a high negative voltage (around $(-)20$ kV), and triggered by an external signal to a gap switch. We fragmented thin $99.6$\% Al$_2$O$_3$ ceramic tubes (SSA-S: Nikkato Corporation) that measured $1.2$ mm in outside diameter, $0.2$ mm thick, and $200$, $150$, or $100$ mm long, with a density of $3.9$ g/cm$^3$. They were sufficiently long that one might regard them as one-dimensional objects. In fact, a few (low discharge energy) cases produced one-dimensional-like fragments that retained the long tube geometry. However, we confirmed that most of the fragments had two-dimensional geometry, i.e., after fragmentation, very few fragments retained the original tube geometry, particularly not those in the small mass range. Therefore, we considered this experiment a two-dimensional one. Owing to the discharge voltage limit of the generator, we could not break larger samples than those used.

A thin ceramic tube was threaded on a copper wire ($0.05$ mm diameter) and the wire was fixed to copper electrodes in a vessel. When a pulsed electric discharge passed through the wire, the copper vaporized and the resulting pressure inside the ceramic tube increased dramatically, causing fragmentation of the ceramic tube. During the experiments, the current and voltage between the electrodes were measured using a current probe (Model 110A: Pearson Electronics) and a high-voltage probe (EP-50K: Pulse Electric Engineering), respectively. After each fragmentation, all of the fragments were collected and their masses were measured using an electronic balance (ER-60A: A{\&}D). We fragmented $47$ ceramic tubes ($21$: $200$ mm-long, $18$: $150$ mm-long, and $8$: $100$ mm-long). The input parameter of this experiment was the discharge voltage, which ranged from $18$ $\sim$ $23$ kV. When we analyzed the fragment statistics, we used the minimum limit of fragment mass $m_{\min}=0.001$ g. This roughly corresponds to the reliable limit of our measurements. Moreover, this value is sufficiently larger than the smallest limit mass of two-dimensionality $0.2^3$(mm$^3$) $\times 3.9 \times 10^{-3}$(g/mm$^3$) $= 3.1 \times 10^{-5}$ g.

\begin{figure}
\scalebox{1.0}[1.0]{\includegraphics{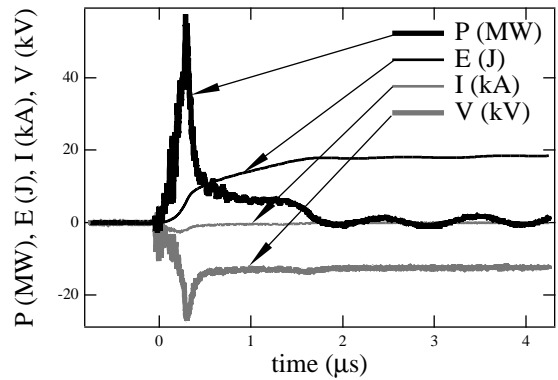}}
\caption{Typical profiles of the current I, voltage V, power P, and discharge energy E as functions of time. A sharp power pulse can be confirmed on an approximately $\mu s$ time scale.}
\label{fig:VIPE}
\end{figure}

First, we show typical profiles of the current, voltage, power, and released energy during a discharge in Fig. \ref{fig:VIPE}. Other experiments showed similar profiles qualitatively. A very sharp power pulse was confirmed. The electrical discharge happened on a ${\mu} s$ time scale, which resulted from the electrical properties of the circuit. This resulted in efficient melting of the copper wire.

\begin{figure}
\scalebox{1.0}[1.0]{\includegraphics{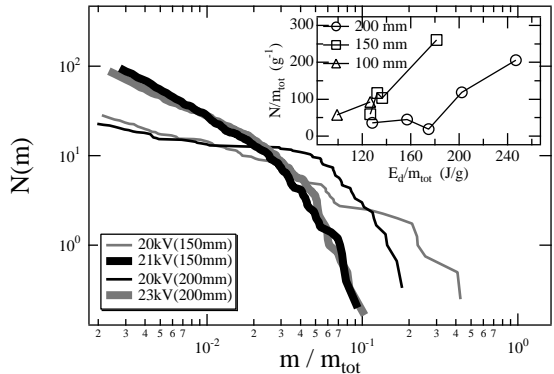}}
\caption{Typical cumulative fragment mass distributions. The distribution curves were obtained from a mean of $4$, $5$, $3$, and $6$ fragmentation events for $20$ kV ($150$ mm), $21$ kV ($150$ mm), $20$ kV ($200$ mm), and $23$ kV ($200$ mm), respectively. The inset shows the relation between the estimated discharge energy and average total fragment number.}
\label{fig:Cum}
\end{figure}

Next, we show merged cumulative fragment mass distributions in Fig. \ref{fig:Cum}. In Fig.\ \ref{fig:Cum}, the $N(m)$ curves are defined as $N(m)= \sum_{i=1}^{r}N_i(m)/r$, where $r$ and $N_i(m)$ are the number of sum-ups and the $i$th cumulative distribution, respectively, i.e., the $N(m)$ curves are averaged. We averaged the results under the same conditions (discharge voltage and sample size). The mass $m$ was normalized to the average total mass $m_{tot}$. For the cases with a high voltage discharge, the $N(m)$ curves show a power-law-like portion. This regime corresponds to the fully fragmented state. By contrast, with low voltage discharge, the $N(m)$ curves appear almost flat in the small mass region. They clearly differ from the fully fragmented regime distributions, and are similar to an integrated log-normal distribution. These features resemble those of impact fragmentation \cite{Katsuragi1,Katsuragi2}. We also show the relation between the estimated discharge energy per unit mass ($E_d / m_{tot}$) and the average total fragment number per unit mass ($N / m_{tot}$) in the inset of Fig.\ \ref{fig:Cum}.
Roughly speaking, the estimated discharge energy shows the extent of the inner pressure produced, and the average total fragment number indicates the degree of fragmentation.

\begin{figure}
\scalebox{1.0}[1.0]{\includegraphics{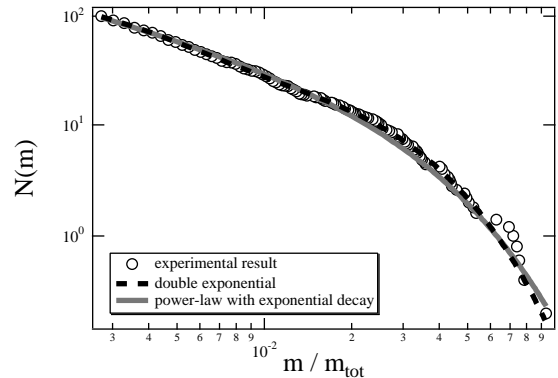}}
\caption{Fitting results using the double exponential and power-law with exponential decay forms. Circles indicate the result for $150$-mm-long tubes with a $21$ kV discharge. Both fittings make good agreement.}
\label{fig:fits}
\end{figure}

We tried to fit the data in the fully fragmented regime using two traditional forms: the double exponential form $N(m) = A_1 \exp(-m/m_1) + A_2 \exp(-m/m_2)$, and the power-law with exponential decay form $N(m)= A_0 m^{-(\tau-1)}\exp(-m/m_0)$. Where $\tau$, $m_0$, $m_1$, $m_2$, $A_0$, $A_1$, and $A_2$ are fitting parameters. The results of fitting for the $150$-mm-long and $21$-kV-discharge case are shown in Fig.\ \ref{fig:fits}. Both function forms agree with the experimental results very well.  Other $N(m)$ curves for the fully fragmented regime were also fitted using both forms. We cannot conclude which form is more appropriate to explain the experiment.

The generic scaling form proposed by {\AA}str\"om et al. (Eq.\ (3) in Ref.\ \cite{Astrom4}) is close to the double exponential form when the penetration depth in the scaling form is small. The penetration depth indicates the strength of crack progression whose dynamics result in a power-law fragment mass distribution \cite{Astrom4}. In our experiment, a small penetration depth is reasonable because the pressure loads were applied uniformly to the two-dimensional (2D) space of the samples. Fragmentation occurred simultaneously throughout 2D space due to this loading condition symmetry. Then, crack nucleation determines the fragment mass distribution rather than the deep crack dynamics. Consequently, the fragment mass distribution approximates the double exponential form due to the Poisson random nucleation of a crack and the dynamics of neighboring cracks that lack a sufficient power-law (i.e., a small penetration depth). In fact, the horizontal sandwich experimental results in Ref.\ \cite{Katsuragi1} can be fitted using {\AA}str\"om et al.'s scaling form when that is close to the double exponential form \cite{Astrom5}. The horizontal sandwich experiment has similar loading condition symmetry to the present explosive experiment.

Conversely, an empirical form of the power-law with exponential decay also results in good agreement. This form has fewer fitting parameters than the double exponential form and is consistent with other experiments, analyses, and simulations performed to date. A weakness of this fitting is the narrow region of the power-law regime in our data. Since the linear part of the log-log plot covers less than one order of magnitude, it might not be sufficient to conclude that the power-law form is valid. Note that the double exponential function can produce a similar portion, as seen in Fig.\ \ref{fig:fits}. We found $\tau-1$ values in the range $0.66-0.76$ for the fully fragmented regime. These values differ from the results of Wittel et al.\ \cite{Wittel1}. There are two free boundaries in the geometry of a tube, whereas an eggshell has only one. To check the boundary effect, we also experimented with two-tone colored tubes. However, we found no significant difference between the regions near the ends and that in the middle of the tubes. While Wittel et al.'s explosion arose from a point source and expanded spherically, we used a linear source and the explosion expanded cylindrically. The curvature of the samples might also have affected the results. These reasons might explain the differences in the results, although we think that the problem remains unresolved.

\begin{figure}
\scalebox{1.0}[1.0]{\includegraphics{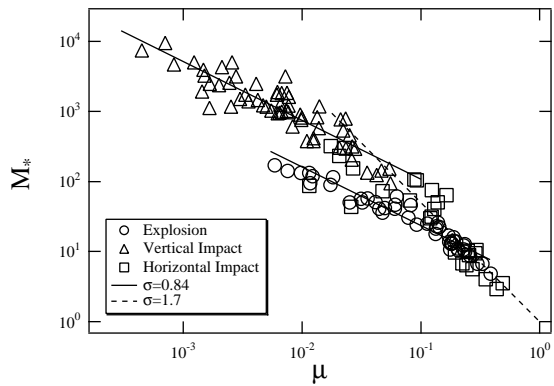}}
\caption{Log-log plot of the weighted mean fragment mass $M_*$ vs. the multiplicity $\mu$. The crossover point $\mu_c$ is not universal for impact and explosive fragmentation. However, the slopes (exponent) appear to be universal.}
\label{fig:M2M1}
\end{figure}

We also examined the behavior of the $k$th order moment $M_k=\sum_m m^k n(m)$ in order to elucidate the critical scaling in the explosive fragmentation. Note that this summation involves the largest fragment mass. Similar to the previous analyses \cite{Katsuragi1,Katsuragi2}, we show the log-log plot of $M_* = M_2 / (M_1 m_{\min})$ vs. $\mu=M_0 m_{\min}/M_1$ in Fig.\ \ref{fig:M2M1}. Figure \ref{fig:M2M1} also shows previous results for impact fragmentation \cite{Katsuragi1,Katsuragi2}, where $M_*$ and $\mu$ are the weighted mean fragment mass and multiplicity, respectively \cite{Katsuragi1,Katsuragi2,Campi1}. When we assume the scaling $M_* \sim \mu^{-\sigma}$, we can confirm the crossover of the scaling exponent $\sigma$ at a certain $\mu_c$ \cite{Katsuragi2}. As Fig.\ \ref{fig:M2M1} shows, a crossover from a small $\sigma$ value to a large one can also occur in explosive fragmentation. Nevertheless, the crossover point $\mu_c$ does not seem to be universal for impact and explosive fragmentation. Although we have conjectured on the universality of $\mu_c$ in Ref.\ \cite{Katsuragi2}, it shows non-universal behavior. However, this is not surprising. The slopes (exponent) $\sigma$ appear to be universal in the two. We think that this is reasonable in terms of critical universality. In general, the critical point is not universal despite the universality of critical exponents. Moreover, this difference in $\mu_c$ might arise from differences in the materials, i.e., glass or ceramic.

\begin{figure}
\scalebox{1.0}[1.0]{\includegraphics{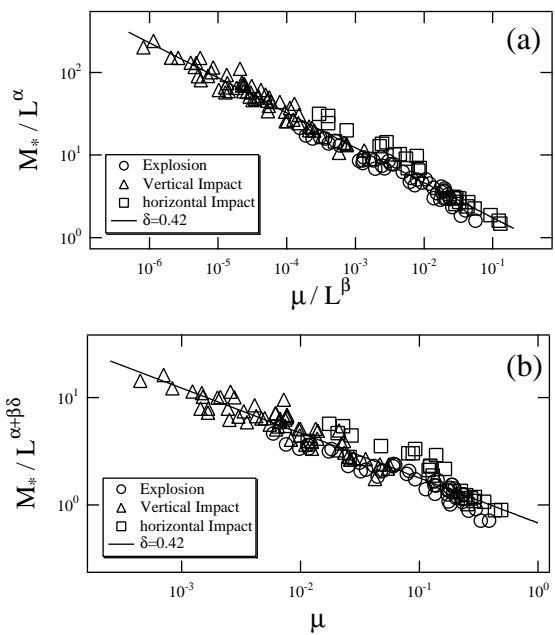}}
\caption{Universal scaling plots of the weighted mean fragment mass. We use the values $\alpha=0.4$ and $\beta=0.7$. We show (a) $M_* / L^{\alpha}$ vs. $\mu / L^{\beta}$, and (b) $M_* / L^{\alpha + \beta \delta}$ as a function of $\mu$.
}
\label{fig:univ}
\end{figure}

We would like to determine a more universal law for fragmentation criticality. In order to improve the scaling, we use the normalized largest fragment mass $L=m_L/m_{\min}$, where $m_L$ is the largest fragment mass. In the finite size system, $L$ plays an important role for scaling analyses. Finally, we assume the universal scaling of the weighted mean fragment mass, as shown in Figs.\ \ref{fig:univ}, which include both impact and explosive results. The scaling form should be written as $M_* / L^{\alpha}=f( \mu / L^{\beta} )$. In Fig.\ \ref{fig:univ}(a), we use the values $\alpha=0.4$ and $\beta=0.7$. From Fig.\ \ref{fig:univ}(a), we deduce the scaling function $f(x) \sim x^{-\delta}$. We obtain the exponent $\delta=0.424 \pm 0.006$ from fitting the data in Fig.\ \ref{fig:univ}(a). Furthermore, this implies $M_* / L^{\alpha+\beta \delta} \sim \mu^{-\delta}$. We show the log-log plot of $M_* / L^{\alpha+\beta \delta}$ as a function of $\mu$ in Fig.\ \ref{fig:univ}(b). The $\delta$ value from Fig.\ \ref{fig:univ}(b) is $\delta=0.42 \pm 0.01$. In either case, the scaling appears convincing globally. We believe that must be a crucial scaling for critical fragmentation. Specifically, the weighted mean fragment mass $M_*$ can be scaled as $L^{0.69}\mu^{-0.42}$ independent of the fragment mass distribution type: log-normal, power-law, or double exponential. The scaling is very robust for small to large imparted energies.

If we assume the binomial multiplicative model with parameter $a=2/3$ \cite{Katsuragi1} for the small $\mu$ regime, we can calculate $\delta_{\mbox{\tiny{bin}}}$ as
\begin{equation}
\delta_{\mbox{\tiny{bin}}} = \frac{\alpha \log a - \log[a^2 + (1-a)^2]}{\log 2 - \beta \log a}.
\label{eq:delta1}
\end{equation}
Substituting $\alpha=0.4$, $\beta=0.7$, and $a=2/3$ into Eq.\ (\ref{eq:delta1}), we obtain $\delta_{\mbox{\tiny{bin}}}=0.44$. This value is close to $\delta=0.42$. By contrast, if we assume the power-law form (with $1<\tau<2$) for the large $\mu$ regime, $M_k$ can be expressed as $M_k \sim \int_{1}^{L}m^k m^{-\tau}dm$. From this notation, we can approximate $M_*/L^{\alpha} \sim (\mu / L^{\beta})^{-\delta_{\mbox{\tiny{pow}}}}$ as $[L^{(3-\tau)}/L^{(2-\tau)}] / [L^{\alpha}] \sim ([L^{(\tau -2)}] / [L^{\beta}])^{-\delta_{\mbox{\tiny{pow}}}}$; therefore,
\begin{equation}
\delta_{\mbox{\tiny{pow}}}=\frac{1-\alpha}{2+\beta-\tau}.
\label{eq:delta3}
\end{equation}
When we substitute $\alpha=0.4$, $\beta=0.7$, and $\tau=1.7$ into Eq.\ (\ref{eq:delta3}), we are led to $\delta_{\mbox{\tiny{pow}}}=0.6$. This value is slightly different from $\delta=0.42$. This suggests the limit of the power-law approximation, or the importance of the exponential form. Determining the origin of the values $\alpha$, $\beta$, and $\delta$ remains a future problem. A theoretical study of this scaling would clarify the fragmentation criticality well.

In conclusion, we carried out a two-dimensional explosive fragmentation experiment using pulsed power. The $N(m)$ curves for the fully fragmented regime could be fitted using the double exponential form or the power-law with exponential decay. Each form has merits and demerits. The weighted mean fragment mass $M_*$ was scaled using the multiplicity $\mu$ and its scaling showed crossover, just as in impact fragmentation. While the scaling exponent $\sigma$ was the same as for impact fragmentation, the crossover point $\mu_c$ differed. Using the normalized largest fragment mass $L$, we obtained universal scaling between the impact (both vertically and horizontally) and explosive fragmentation data. This scaling also appears independent of the material broken.

We appreciate J. A. {\AA}str\"om for fruitful discussion. 
This research was supported by the Japanese Ministry of Education, Culture, Sports, Science and Technology, Grant-in-Aid for Young Scientists, No.\ 16740206.

\end{document}